\documentclass[prb,twocolumn,showpacs,amsmath,amssymb,balancelastpage,citeautoscript]{revtex4-1}

\pdfoutput=1

\bibpunct{[}{]}{,}{n}{}{}
\makeatletter
\def\NAT@def@citea{\def\@citea{\NAT@separator}}
\makeatother

\usepackage{dcolumn} 

\usepackage{xcolor,graphicx,mathtools,bm}

\definecolor{darkblue}{RGB}{0,0,150}
\definecolor{nightblue}{RGB}{0,0,100}

\usepackage[
colorlinks,
citecolor=darkblue,
linkcolor=darkblue,
urlcolor=nightblue]{hyperref}

\usepackage[english]{babel}
\usepackage[babel,kerning=true,spacing=true]{microtype}

\def\alf{\alpha}
\def\eps{\epsilon}
\def\gam{\gamma}
\def\Gam{\Gamma}
\def\lam{\lambda}
\def\Lam{\Lambda}
\def\om{\omega}
\def\Om{\Omega}
\def\sg{\sigma}
\def\Sg{\Sigma}

\def\bk{{\bf k}}

\def\bq{{\bf q}}

\def\b0{{\bf 0}}

\def\cP{{\cal P}}

\def\lra{\leftrightarrow}

\def\sgn{{\rm sgn}}
\def\det{{\rm det}}

\begin{document}

\title{Fermion loops and improved power-counting in two-dimensional\\ critical metals with singular forward scattering}

\author{Tobias Holder}
\affiliation{Max-Planck-Institute for Solid State Research,
 D-70569 Stuttgart, Germany}
\author{Walter Metzner}
\affiliation{Max-Planck-Institute for Solid State Research,
 D-70569 Stuttgart, Germany}

\date{\today}

\begin{abstract}
We analyze general properties of the perturbation expansion for two-dimensional quantum critical metals with singular forward scattering, such as metals at an Ising nematic quantum critical point and metals coupled to a $U(1)$ gauge field.
We derive asymptotic properties of fermion loops appearing as subdiagrams of the contributing Feynman diagrams -- for large and small momenta.
Substantial cancellations are found in important scaling limits, which reduce the degree of divergence of Feynman diagrams with boson legs.
Implementing these cancellations we obtain improved power-counting estimates that yield the true degree of divergence.
In particular, we find that perturbative contributions to the boson self-energy are generally ultraviolet convergent for a dynamical critical exponent $z<3$, and divergent beyond three-loop order for $z \geq 3$.
\end{abstract}

\maketitle


\section{Introduction}

The low-energy behavior of interacting fermion systems can be strongly influenced by collective bosonic degrees of freedom such as critical order parameter fluctuations or emergent gauge fields.
In metallic systems, the scattering of electrons by gapless bosons destroys Landau quasi-particles and thus leads to a breakdown of Fermi liquid theory~\cite{loehneysen07}.
These effects are particularly pronounced in two-dimensional systems, and they have therefore frequently been discussed in the context of layered compounds such as cuprate- or iron-based high-temperature superconductors.
In this paper we focus on systems in which the critical bosons carry a small momentum and couple to the electron charge, leading thus to a singular but spin-conserving {\em forward scattering}\/ of electrons.
This class of systems comprises electron liquids coupled to emergent $U(1)$ gauge fields~\cite{lee06}, as well as quantum critical metals at the onset of nematic or other translation-invariant charge order~\cite{oganesyan01, metzner03, metlitski10}.

The breakdown of Fermi liquid behavior due to singular forward scattering is revealed already by first-order perturbation theory.
One-loop results for the boson and fermion self-energies in two-dimensional metals were first derived for fermions coupled to a $U(1)$ gauge field~\cite{lee89}, and later for systems at a nematic quantum critical point (QCP)~\cite{oganesyan01,metzner03}.
The bosons receive a Landau damping term generated by fermionic particle-hole excitations, and the one-loop fermion self-energy scales as $|\om|^{2/3}$ at low frequencies $\om$.
This self-energy dominates over the linear frequency term in the fermion propagator and thus implies a destruction of Landau quasi-particles.
The main contributions to the fermion self-energy at a fixed Fermi momentum $\bk_F$ come from particle-hole excitations near the same $\bk_F$ and its antipode $-\bk_F$, with small momentum transfers $\bq$ tangential to the Fermi surface and excitations energies of the order $|\bq|^3$.
The coupled fermion-boson theory is scale-invariant with a dynamical exponent $z=3$ at the one-loop level~\cite{polchinski94}.

An early analysis of two-loop corrections did not reveal any qualitative changes of the boson and fermion propagators~\cite{altshuler94}.
The loop expansion was expected to be controlled by the inverse fermion flavor number $N_f$, as usual~\cite{polchinski94}.
However, the whole case was opened again when S.-S.\ Lee~\cite{lee09} discovered that the naive $1/N_f$ expansion breaks down and Feynman diagrams of any loop order contribute even in the limit $N_f \to \infty$.
Subsequently, Metlitski and Sachdev~\cite{metlitski10} constructed a general scaling theory for two-dimensional metals at a nematic QCP and the formally similar $U(1)$ gauge field problem.
Symmetry constraints restrict the theory such that only two independent anomalous scaling exponents are possible: an anomalous dimension of the fermion fields $\eta_f$ and an anomalous dynamical scaling exponent $z \neq 3$.
A small contribution to $\eta_f$ was indeed found at {\em three-loop}\/ order,
while the dynamical exponent remained unrenormalized at $z=3$ at that loop level~\cite{metlitski10}.
Most recently it was shown that the absence of renormalizations of $z$ up to three-loop order has special reasons that do not apply to all higher-order contributions. A divergence leading to anomalous dynamical scaling was found in a class of {\em four-loop} diagrams~\cite{holder15}.
Surprisingly, the divergence turned out to be stronger than logarithmic, indicating an anomaly or instability whose nature has not yet been clarified.

Naive power-counting frequently overestimates the actual degree of divergence of perturbative contributions. Cancellations may occur for single Feynman diagrams, due to oscillating integrands, and also between distinct Feynman diagrams.
In this context, the scaling behavior of {\em fermion loops}\/ plays a particularly important role. Most Feynman diagrams contain fermion loops that are connected to each other and/or to open fermion lines by bosons.
All hitherto observed cancellations are related to cancellations within a loop or between loops with permuted vertices.

In this paper we analyze the asymptotic behavior of fermion loops (with $N$ vertices) and sums of loops with permuted vertices in the scaling limit that applies to fermions coupled to bosons in two dimensions as described above.
We perform the analysis for a generalized class of theories with dynamical exponents $z > 2$, as originally introduced by Nayak and Wilczek~\cite{nayak94}.
Systematic cancellations are obtained if one or several of the external loop momenta are much smaller than the others.
As an application, we can assess the degree of divergence of several classes of Feynman diagrams without embarking on a tedious specific calculation of the corresponding integrals.

The paper is structured as follows.
In Sec.~II we introduce the low-energy effective quantum field theory for the systems described above, and we provide the precise definition of the fermion loops.
The result of the elementary integration over the momentum in the loop is presented in Sec.~III.
The asymptotic behavior of fermion loops with three vertices is analyzed in Sec.~IV. An important application is a deeper understanding of the mechanism leading to the absence of a renormalization of the dynamical exponent $z$ in the three-loop calculation by Metlitski and Sachdev~\cite{metlitski10}.
Asymptotic properties of fermion loops with an arbitrary number of vertices $N$ are derived in Sec.~V. In particular, it will be clarified to what extent cancellations obtained at three-loop order can be expected also at higher orders.
In Sec.~VI we provide several examples of improved power-counting.
We finally summarize the main results in Sec.~VII.


\section{Field theory and fermion loops}

The low-energy behavior of two-dimensional fermion systems coupled to critical bosons with small momenta can be described by an effective field theory whose fermionic states are restricted to two Fermi ``patches'' near fixed Fermi momenta $\bk_F$ and $-\bk_F$~\cite{lee09,metlitski10}.
Choosing momentum variables such that the $x$-component is perpendicular and the $y$-component tangential to the Fermi surface at $\pm \bk_F$, the Lagrangian of the field theory can be written in the form~\cite{metlitski10}
\begin{align} \label{lagrangian}
 L &= \sum_{s=\pm} \psi^\dagger_s
 \left( \partial_\tau -
 is \partial_x - \partial^2_y \right) \psi_s \notag\\
 & \quad -\sum_{s=\pm} g_s\phi\psi^\dagger_s\psi_s
 - \frac{N_f}{2e^2}(\partial_y\phi)^2 .
\end{align}
Here $\phi$ is a bosonic scalar field, while $\psi_{\pm}$, $\psi_{\pm}^{\dagger}$ are fermionic fields with $N_f$ flavor components corresponding to states with momenta on the the two patches near $\pm\bk_F$. 
In the U(1)-gauge field problem, $\phi$ is the transverse gauge field and $g_+ = -g_-$. For the Ising nematic QCP, $\phi$ is the order parameter field and $g_+ = g_-$. In both cases the physical flavor number is $N_f = 2$.
The derivatives are with respect to real space and imaginary time variables.
Several numerical prefactors have been absorbed by a rescaling of fields and space coordinates. In particular $|g_s| = 1$.
The gradient term in $x$-direction $(\partial_x \phi)^2$ is irrelevant under the scaling that emerges on one-loop level, and has therefore been discarded in the Lagrangian.

In random-phase approximation (RPA), which corresponds to a one-loop calculation of the boson and fermion self-energies, the boson and fermion propagators are obtained as~\cite{lee89}
\begin{align}
 D^{-1}(q) &= 
 N_f \left( \frac{q_y^2}{e^2} + 
 \gam \frac{|q_0|}{|q_y|} \right), \\
 G^{-1}_s(k) &=
 s k_x + k_y^2 - i \frac{\kappa}{N_f} \frac{k_0}{|k_0|^{1/3}} ,
\end{align}
where $q = (q_0,\bq)$ and $k = (k_0,\bk)$ contain momentum and frequency variables, and $\gam$ and $\kappa$ are positive constants.
These propagators solve the RPA equations also self-consistently~\cite{polchinski94}.
The linear frequency term in the fermion propagator is subleading compared to the fermion self-energy and has therefore been discarded.
Note that $D(q)$ does not depend on $q_x$.

In the following we consider a generalization of the theory where $q_y^2$ is replaced by $q_y^{z-1}$ in the boson propagator, such that
\begin{equation}
 D^{-1}(q) = 
 N_f \left( \frac{|q_y|^{z-1}}{e^2} + \gam \frac{|q_0|}{|q_y|} \right) \, .
\end{equation}
This generalization was introduced by Nayak and Wilczek~\cite{nayak94} for the sake of an expansion in $\eps = z-2$. More recently, it was used to define a manageable large-$N_f$ limit of the theory~\cite{mross10}.
For any $z>2$, the one-loop fermion propagator then assumes the generalized form
\begin{equation} \label{Gk}
 G_s(k) = \frac{1}
 {sk_x + k_y^2 - i\{k_0\}} \, ,
\end{equation}
where 
\begin{equation}
 \{k_0\} = \frac{\kappa}{N_f} \frac{k_0}{|k_0|^{\alf}} \quad
 \mbox{with} \quad \alf = 1- \frac{2}{z} \, .
\end{equation}

The fermion propagator scales homogeneously as $G(k) \mapsto \lam^{-2} G(k)$ under the scaling of momentum and frequency variables
\begin{equation} \label{scaling}
 k_y \mapsto \lam k_y , \quad
 k_x \mapsto \lam^2 k_x , \quad
 k_0 \mapsto \lam^z k_0 .
\end{equation}
The boson propagator scales as $D(q) \mapsto \lam^{1-z} D(q)$ under the analogous scaling of $q_0$, $q_x$, and $q_y$.
For $z=3$ we recover $\alf = 1/3$, and $D(q)$ scales with the same power $\lam^{-2}$ as $G(k)$.

Higher-order contributions (beyond one-loop) to the fermion and boson self-energies are analyzed in an expansion around the one-loop fixed point, that is, by computing Feynman diagrams with one-loop propagators~\cite{altshuler94,lee09,metlitski10}.
Note that this procedure can be viewed as a simple shift of the expansion point, where the one-loop self-energy is added to the ``bare'' part of the action and subtracted as a counterterm from the interaction part.

The computation and asymptotic analysis of higher-order contributions is done most efficiently by evaluating first the fermion loops in the Feynman diagrams, and then the remaining bosonic momentum integrations.
The {\em $N$-point fermion loop}\/ on patch $s$ is defined as the integrated product of $N$ fermion propagators,
\begin{eqnarray} \label{PiN}
 \Pi_{N,s}(q_1,\dots,q_N) &=& I_{N,s}(p_1,\dots,p_N) \nonumber \\ 
 &=& \int \frac{dk_0}{2\pi} \int \frac{d^2\bk}{(2\pi)^2} 
 \prod_{i=1}^N G_s(k-p_i) \, . \hskip 5mm
\end{eqnarray}
The variables $p_i$ and $q_i$ are related by
\begin{eqnarray} \label{pq}
 q_i &=& p_{i+1} - p_i \quad \mbox{for} \quad i = 1,\dots,N-1 
 \nonumber \\
 q_N &=& p_1 - p_N \, .
\end{eqnarray}
Note that $I_{N,s}(p_1,\dots,p_N)$ is invariant under a global shift of momenta $p_i \mapsto p_i + q$, and $q_1 + \dots + q_N = 0$ due to energy and momentum conservation.
$\Pi_{N,s}(q_1,\dots,q_N)$ can be represented graphically as a fermion loop, as shown in Fig.~1.
\begin{figure}[t]
\centering
\includegraphics[width=7cm]{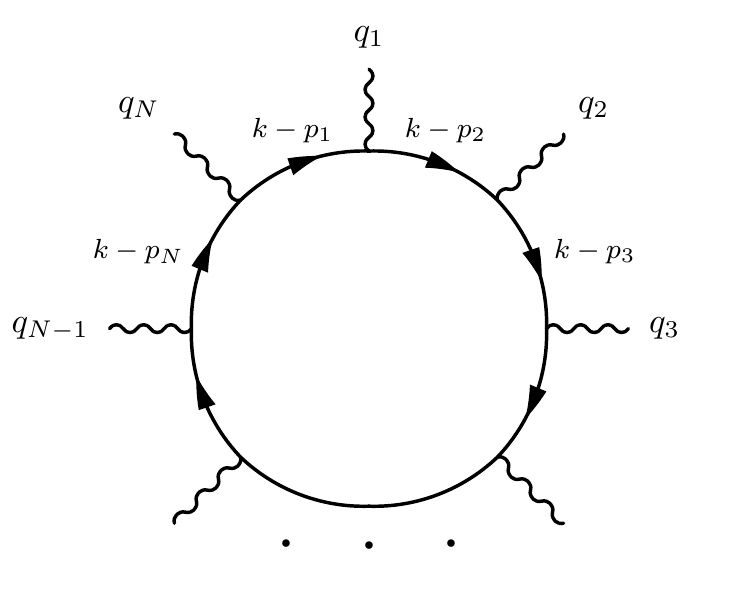}
\caption{Graphical representation of the $N$-point fermion loop.}
\end{figure}

The sum of Feynman diagrams contributing at a certain order in the loop expansion can be written in terms of {\em symmetrized}\/ fermion loops,
\begin{equation}
 \Pi_{N,s}^{\rm sym}(q_1,\dots,q_N) = \frac{1}{N}
 \sum_{\cP} \Pi_{N,s}(\cP q_1,\dots,\cP q_N) \; ,
\end{equation}
with a sum over all permutations of $q_1,\dots,q_N$.
The prefactor $1/N$ compensates the $N$-fold multiplicity arising from cyclic permutations.
In various limits the symmetrized loops are much smaller than the unsymmetrized contributions, due to systematic cancellations.
Strong cancellations between contributions to symmetrized loops are well-known in one-dimensional Luttinger liquids~\cite{dzyaloshinskii74}, and have also been established for two-dimensional Fermi liquids~\cite{neumayr98, kopper01}.

In the remainder of our paper we evaluate the $N$-point loops and analyze their asymptotic behavior in various important limits relevant to quantum critical metals, from which we can derive improved power-counting estimates of Feynman diagrams.
To simplify the notation, we specify to fermion loops defined on the Fermi patch with $s=+$ and drop the patch index.
Results on the other patch ($s=-$) can be obtained from those for $s=+$ by simply switching the sign of the $x$-component of all momenta.
Furthermore, we absorb the prefactor $\kappa/N_f$ in the fermion propagator by a rescaling of the frequency variable. Hence, $\{ k_0 \}$ stands for $k_0/|k_0|^\alf$ in the following sections.


\section{Integration of fermion loops}

The fermion loop as defined in Eq.~(\ref{PiN}) involves a momentum and a frequency integration. The momentum integral can be performed analytically by using the residue theorem (see Appendix A), yielding
\begin{eqnarray} \label{IN}
 I_N(p_1,\dots,p_N) &=& \sum_{i<j}
 \int_{p_{j0}}^{p_{i0}} \frac{dk_0}{4\pi} \,
 \Theta\Big(\frac{p_{i0} - p_{j0}}{p_{iy} - p_{jy}}\Big) \nonumber \\
 &\times& (p_{iy} - p_{jy})^{N-3} \prod_{l \neq i,j} J_{ijl}(k_0)
\end{eqnarray}
for any $N \geq 3$. The indices $i,j,l$ run from 1 to $N$ with restrictions as indicated, $\Theta$ is the step function, and
\begin{equation} \label{Jijl}
 J_{ijl}(k_0) = \frac{1}{D_{ijl} + F_{ijl} + i\Om_{ijl}(k_0)} \; ,
\end{equation}
with
\begin{eqnarray}
 D_{ijl} &=& p_{ix}(p_{ly} - p_{jy}) + {\rm cyc} \, , \label{Dijl} \\
 F_{ijl} &=& (p_{jy} - p_{iy})(p_{ly} - p_{jy})(p_{iy} - p_{ly}) \, , 
 \label{Fijl} \\[2mm]
 \Om_{ijl}(k_0) &=&
 \{k_0 - p_{i0}\} (p_{ly} - p_{jy}) +
 {\rm cyc} \, , \label{Omijl}
\end{eqnarray}
where ``cyc'' denotes cyclic permutations of the indices $i,j,l$.
Note that the remaining frequency integration is limited to finite intervals and hence convergent. The fermion loop integral does not require any ultraviolet regularization.
In the static limit $q_{i0} = 0$, all frequency variables $p_{i0}$ are equal, such that $I_N(p_1,\dots,p_N) = 0$.

The quantities $D_{ijl}$, $F_{ijl}$, and $\Om_{ijl}(k_0)$ are antisymmetric in the indices $i,j,l$. Hence, $J_{ijl}(k_0)$ is also antisymmetric in its indices.
Using this antisymmetry, the three-point loop can be written in the particularly simple form
\begin{eqnarray} \label{I3}
 I_3(p_1,p_2,p_3) &=& \sum_{(i,j) = (1,2),(2,3),(3,1)}
 \int_{p_{j0}}^{p_{i0}} \frac{dk_0}{4\pi} \,
 \Theta\Big(\frac{p_{i0} - p_{j0}}{p_{iy} - p_{jy}}\Big) \nonumber \\
 &\times& J_{123}(k_0) \, .
\end{eqnarray}
Under the scaling Eq.~(\ref{scaling}), the quantities $D_{ijl}$, $F_{ijl}$, and $\Om_{ijl}(k_0)$ scale homogeneously as
\begin{equation}
 D_{ijl} \mapsto \lam^3 D_{ijl} , \;
 F_{ijl} \mapsto \lam^3 F_{ijl} , \;
 \Om_{ijl}(k_0) \mapsto \lam^3 \Om_{ijl}(k_0) \, .
\end{equation}
Hence $J_{ijl}(k_0) \mapsto \lam^{-3} J_{ijl}(k_0)$ and
\begin{equation}
 I_N(p_1,\dots,p_N) \mapsto \lam^{3+z-2N} I_N(p_1,\dots,p_N) \, ,
\end{equation}
in agreement with direct power-counting applied to the definition of $I_N(p_1,\dots,p_N)$ as a loop-integral.
The symmetrized fermion loops $\Pi_N^{\rm sym}(q_1,\dots,q_N)$ exhibit the same scaling behavior. As a consequence, effective boson interactions with amplitudes proportional to $\Pi_N^{\rm sym}(q_1,\dots,q_N)$ are marginal for all $N$~\cite{thier11}.

In the special case $\alf = 0$, corresponding to a bare fermion progagator, the $k_0$-dependence in $\Om_{ijl}(k_0)$ cancels in the cyclic sum, such that $\Om_{ijl}(k_0) = \Om_{ijl} = p_{i0} (p_{jy} - p_{ly}) + {\rm cyc}$.
The frequency integration in Eq.~(\ref{IN}) then becomes trivial, yielding
\begin{eqnarray} \label{IN0int}
 I_N^0(p_1,\dots,p_N) &=& \frac{1}{4\pi} \sum_{i<j}
 \Theta\Big(\frac{p_{i0} - p_{j0}}{p_{iy} - p_{jy}}\Big)
 (p_{i0} - p_{j0}) \nonumber \\
 &\times& (p_{iy} - p_{jy})^{N-3} \prod_{l \neq i,j} J_{ijl} \; .
\end{eqnarray}
In particular, the three-point loop with bare propagators is obtained as
\begin{eqnarray} \label{I30}
 I_3^0(p_1,p_2,p_3) &=& \frac{1}{4\pi} \sum_{(i,j) = (1,2),(2,3),(3,1)}
 \Theta\Big( \frac{p_{i0} - p_{j0}}{p_{iy} - p_{jy}} \Big) \nonumber \\
 &\times& (p_{i0} - p_{j0}) \, J_{123} \, .
\end{eqnarray}
In the infrared scaling limit $p_{i0} \mapsto \lam^z p_{i0}$, $p_{ix} \mapsto \lam^2 p_{ix}$, $p_{iy} \mapsto \lam p_{iy}$ with $z>2$ and $\lam \to 0$, the term $\Om_{123}$ is subleading compared to $D_{123}$ and $F_{123}$ in the denominator of $J_{123}$. In that limit, Eq.~(\ref{I30}) is consistent with the result for the $3$-point loop with a full Fermi surface integral obtained previously in Ref.~\onlinecite{thier11}.


\section{Asymptotic behavior of three-point loop}

In this section we derive asymptotic scaling properties of the three-point loop. In particular, we analyze the behavior in a limit that determines the ultraviolet behavior of the Aslamasov-Larkin-type diagrams depicted in Fig.~2.
These diagrams have been computed explicitly in the static limit (external $q_0=0$) by Metlitski and Sachdev~\cite{metlitski10}.
\begin{figure}[b]
\centering
\includegraphics[width=8cm]{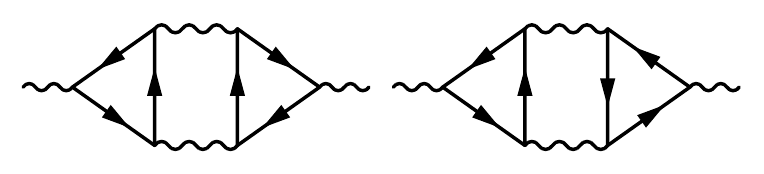}
\caption{Aslamasov-Larkin type diagrams contributing to the bosonic self-energy in a coupled theory involving fermions (solid lines) and bosons (wiggly lines).}
\end{figure}

Denoting the external frequency-momentum variable in the Aslamasov-Larkin diagrams by $q$, and the bosonic loop integration variable by $q'$, the contributing fermion loops are $\Pi_3(q,q',-q-q') = I_3(0,q,q+q')$ and $\Pi_3(q,-q-q',q') = I_3(0,q,-q') = I_3(0,-q,-q-q')$.
In the static limit $q_0 = 0$, Eq.~(\ref{I3}) yields
\begin{eqnarray}
 I_3(0,q,q+q') &\stackrel{q_0=0}{=}& \int_0^{q'_0} \frac{dk_0}{4\pi}
 \left[ \Theta \Big( \frac{q'_0}{q_y + q'_y} \Big) -
 \Theta \Big( \frac{q'_0}{q'_y} \Big) \right] \nonumber \\
 &\times& J_{123}(k_0) \, ,
\end{eqnarray}
with
\begin{eqnarray}
 J_{123}(k_0) &=& \big[ q'_x q_y - q_x q'_y - q_y q'_y (q_y + q'_y)
 \nonumber \\
 && + i\left( \{k_0\} - \{k_0-q'_0\} \right) q_y \big]^{-1} \, .
\end{eqnarray}
Hence, $I_3(0,q,q+q')$ with $q_0 = 0$ is non-zero only if $|q'_y| < |q_y|$ and $\sgn(q'_y) = - \sgn(q_y)$. This {\em kinematic constraint} restricts the $q'_y$ integral to an interval of length $|q_y|$.

Now we can assess the ultraviolet (UV) behavior of the $q'$-integral in the Aslamasov-Larkin diagrams. We introduce an ultraviolet cutoff $\Lam$ such that $|q'_0| < \Lam^z$, $|q'_x| < \Lam^2$, and $|q'_y| < \Lam$. However, for fixed $q_y$, the $q'_y$-integral is effectively restricted by $q_y$. With this restriction, one has
\begin{equation}
 D_{123} \sim q_y \Lam^2, \quad F_{123} \sim q_y^3, \quad
 \Om_{123}(k_0) \sim q_y \Lam^2 \, ,
\end{equation}
and thus $J_{123}(k_0) \sim (q_y \Lam^2)^{-1}$, such that the UV behavior of the three-point loop is
\begin{equation}
 I_3(0,q,q+q') \sim \frac{\Lam^{z-2}}{q_y} \, .
\end{equation}
The Aslamasov-Larkin diagrams contain two three-point fermion loops and two boson propagators. For $q'_y \sim q_y$ and $q'_0 \sim \Lam^z$ the latter scale as $q_y/\Lam^z$. The integration measure scales as $\Lam^z \Lam^2 q_y = \Lam^{z+2} q_y$. Hence, the integral diverges in the UV limit as $q_y \Lam^{z-2}$. This agrees with the result obtained in a more explicit calculation for $z=3$ by Metlitski and Sachdev~\cite{metlitski10}.
Note that the above contribution is obtained only if the two fermion loops are integrated over antipodal Fermi patches. For loops on the same Fermi patch, the $q'_x$-integral vanishes since the integration contour can be closed without encircling any poles in that case~\cite{lee09,metlitski10}.

The sum of all Aslamasov-Larkin diagrams can be expressed by symmetrized loops.
The symmetrized three-point loop can be written as
\begin{equation}
 \Pi_3^{\rm sym}(q_1,q_2,q_3) = I_3(p_1,p_2,p_3) + I_3(-p_1,-p_2,-p_3) \, .
\end{equation}
Under the inversion $p_i \mapsto -p_i$ and $k_0 \mapsto -k_0$, the quantities $D_{ijl}$, $F_{ijl}$ and $\Om_{ijl}(k_0)$ transform as
\begin{equation}
 D_{ijl} \mapsto D_{ijl} , \quad F_{ijl} \mapsto - F_{ijl} , \quad
 \Om_{ijl}(k_0) \mapsto \Om_{ijl}(k_0) \, .
\end{equation}
The symmetrized three-point loop contributing to the Aslamasov-Larkin diagrams in the static limit can thus be written as
\begin{eqnarray}
 I_3^{\rm sym}(0,q,q+q') \stackrel{q_0=0}{=} \int_0^{q'_0} \frac{dk_0}{4\pi}
 \left[ \Theta \Big( \frac{q'_0}{q_y + q'_y} \Big) -
 \Theta \Big( \frac{q'_0}{q'_y} \Big) \right] \nonumber \\
 \times \! \left[ 
 \frac{1}{D_{123} \!+\! F_{123} \!+\! i\Om_{123}(k_0)}
 - \frac{1}{D_{123} \!-\! F_{123} \!+\! i\Om_{123}(k_0)} \right] \! . \hskip 3mm
\end{eqnarray}
In the ultraviolet limit the leading terms in the last set of brackets cancel, leaving a much smaller contribution of order $q_y/\Lam^4$, such that the symmetrized three-point loop scales as
\begin{equation}
 I_3^{\rm sym}(0,q,q+q') \sim q_y \Lam^{z-4} \, .
\end{equation}
This is by a factor $(q_y/\Lam)^2$ smaller than the unsymmetrized three-point loop, that is, the degree of UV divergence has been reduced by two upon symmetrization.
The sum over the two inequivalent Aslamasov-Larkin diagrams is thus ultraviolet-finite at least for $z<4$, as shown previously for $z=3$ by Metlitski and Sachdev~\cite{metlitski10}.


\section{Asymptotic properties of $N$-point loop}

We have seen that the three-point loop is cut off by a kinematic constraint at large momenta, if one of the external momenta stays fixed, and cancellations occur upon symmetrization.
In this section we investigate to what extent these properties can be generalized to $N$-point loops. To this end, we first rewrite $I_N(p_1,\dots,p_N)$ in a form which is more convenient for an asymptotic analysis.


\subsection{Reduction formula for $N$-point loop}

Feldman et al.~\cite{feldman98} derived a {\em reduction formula} expressing the $N$-point loop with bare propagators and a quadratic dispersion relation as a linear combination of three-point loops,
\begin{equation} \label{IN0red}
 I_N^0(p_1,\dots,p_N) = \sum_{i<j<l}
 \Big[ \prod_{\nu \neq i,j,l} f_{ijl,\nu}^{-1} \Big] \,
 I_3^0(p_i,p_j,p_l) \, ,
\end{equation}
where $f_{ijl,\nu}$ is a rational function of $p_i,p_j,p_l,p_{\nu}$~\cite{neumayr99}.

For the $N$-point loop with non-Fermi liquid propagators of the form Eq.~(\ref{Gk}), a similar formula can be derived. In Appendix B we show that $I_N(p_1,\dots,p_N)$ as given by Eq.~(\ref{IN}) can be expressed as
\begin{eqnarray} \label{INred}
 I_N(p_1,\dots,p_N) &=& \sum_{i<j<l}
 \int_{p_{j0}}^{p_{i0}} \frac{dk_0}{4\pi}
 \Theta\Big( \frac{p_{i0}-p_{j0}}{p_{iy}-p_{jy}} \Big) \nonumber \\
 &\times& \Big[ \prod_{\nu \neq i,j,l} f_{ijl,\nu}^{-1}(k_0) \Big]
 J_{ijl}(k_0) + {\rm cyc} , \hskip 5mm
\end{eqnarray}
where
\begin{equation} \label{fk0}
 f_{ijl,\nu}(k_0) = 
 \frac{D_{ijl} J_{ij\nu}^{-1}(k_0) - D_{ij\nu} J_{ijl}^{-1}(k_0)}
 {D_{ijl} (p_{iy} - p_{jy})} \, ,
\end{equation}
with $J_{ijl}(k_0)$ and $D_{ijl}$ from Eqs.~(\ref{Jijl}) and (\ref{Dijl}), respectively. ``cyc'' denotes cyclic permutations of $i,j,l$.
Expressing $J_{ijl}(k_0)$ and $D_{ijl}$ explicitly in terms of momenta and frequencies, one finds that the product $D_{ijl} f_{ijl,\nu}(k_0)$ is an antisymmetric polynomial of the form
\begin{eqnarray}
 D_{ijl} f_{ijl,\nu}(k_0) &=& 
 \left| \begin{array}{llll}
 p_{ix} & p_{jx} & p_{lx} & p_{\nu x} \\
 p_{iy} & p_{jy} & p_{ly} & p_{\nu y} \\
 p_{iy}^2 & p_{jy}^2 & p_{ly}^2 & p_{\nu y}^2 \\
 1 & 1 & 1 & 1 \end{array} \right| \nonumber \\ 
 &-& i \left| \begin{array}{cccc}
 p_{ix} & p_{jx} & p_{lx} & p_{\nu x} \\
 p_{iy} & p_{jy} & p_{ly} & p_{\nu y} \\
 \{k_{i0} \} & \{k_{j0} \} & \{k_{l0} \} & \{ k_{\nu 0} \} \\
 1 & 1 & 1 & 1
 \end{array} \right| , \hskip 5mm
\end{eqnarray}
with $k_{i0} = k_0 - p_{i0}$. Hence, $D_{ijl} f_{ijl,\nu}(k_0)$ is totally antisymmetric in all four indices. Since $D_{ijl}$ is also antisymmetric, it follows that $f_{ijl,\nu}$ is invariant under permutations of $i,j,l$.

For loops constructed with bare propagators $G_0$, the quantities $J_{ijl}$ and $f_{ijl,\nu}$ are independent of $k_0$. The frequency integration in Eq.~(\ref{INred}) can then be carried out and one recovers the reduction formula for $I_N^0(p_1,\dots,p_N)$ in the form Eq.~(\ref{IN0red}).


\subsection{Kinematic constraint for $N$-point loop}

We now analyze the asymptotic behavior of an $N$-point loop traversed by a single large momentum. More precisely, we assume that two of the external momenta, say $q'$ and $q'' \approx -q'$ are large and almost antiparallel, while all the other momenta are kept finite, that is, relatively small.
This implies that the momenta $p_i$ fall in two groups, where momenta within a group are close together, while the distance between momenta in different groups is large (see Fig.~3). Choosing $p_1 = 0$, the momenta in the group containing $p_1$ are all close to zero, that is, relatively small.
\begin{figure}[b]
\centering
\includegraphics[width=5cm]{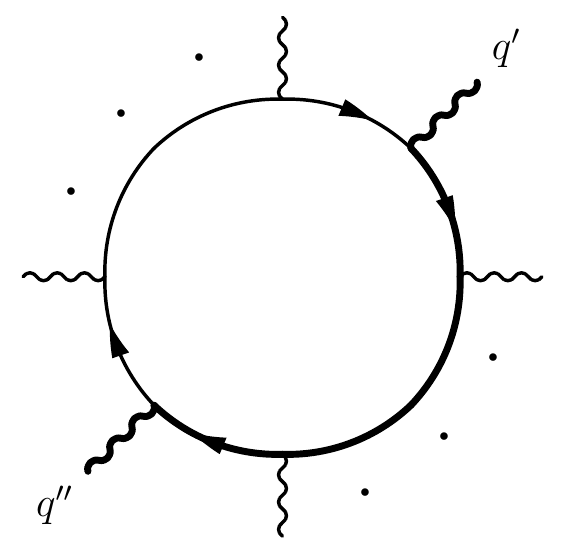}
\caption{$N$-point loop with two large momenta $q'$ and $q'' \approx - q'$. Momenta $p_i$ on the bold lines are large, those on thin lines (relatively) small.}
\end{figure}

Using the generalized reduction formula (\ref{INred}), we can show that the frequency integrations are effectively restricted to small intervals, although the differences $p_{i0} - p_{j0}$ may be large.
To this end, let us fix some $i,j,l$ and write out the cyclic sum:
\begin{eqnarray} \label{cycle1}
 && \int_{p_{j0}}^{p_{i0}} \frac{dk_0}{4\pi} 
 \Theta\Big( \frac{p_{i0} - p_{j0}}{p_{iy} - p_{jy}} \Big) g(k_0) \nonumber \\
 &+& \int_{p_{l0}}^{p_{j0}} \frac{dk_0}{4\pi} 
 \Theta\Big( \frac{p_{j0} - p_{l0}}{p_{jy} - p_{ly}} \Big) g(k_0) \nonumber \\
 &+& \int_{p_{i0}}^{p_{l0}} \frac{dk_0}{4\pi} 
 \Theta\Big( \frac{p_{l0} - p_{i0}}{p_{ly} - p_{iy}} \Big) g(k_0) \, ,
\end{eqnarray}
where $g(k_0) =
 \Big[ \prod_{\nu \neq i,j,l} f_{ijl,\nu}^{-1}(k_0) \Big] J_{ijl}(k_0)$.
Note that the latter function is invariant under cyclic permutations of $i,j,l$.

If $p_i$, $p_j$, $p_l$ are in the same group of small or large momenta, all frequency integrations in (\ref{cycle1}) are obviously limited to small intervals. Now assume that they are in two distinct groups, say $p_i$ is large and the others are small. The argument of the first and third step function in (\ref{cycle1}) is then dominated by $p_i$, leaving $\Theta(p_{i0}/p_{iy})$ in both cases, and the whole expression can be simplified to
\begin{equation} \label{cycle2}
 \int_{p_{j0}}^{p_{l0}} \frac{dk_0}{4\pi} 
 \Theta\Big( \frac{p_{i0}}{p_{iy}} \Big) g(k_0) +
 \int_{p_{l0}}^{p_{j0}} \frac{dk_0}{4\pi} 
 \Theta\Big( \frac{p_{j0} - p_{l0}}{p_{jy} - p_{ly}} \Big) g(k_0) \, .
\end{equation}
The frequency integration is thus effectively restricted to the small interval between $p_{j0}$ and $p_{l0}$. The large contributions from integrations from $p_{j0}$ to $p_{i0}$ and from $p_{i0}$ to $p_{l0}$ cancel in the cyclic sum.

In naive power-counting for the $N$-point loop, the frequency integration yields a factor $\Lam^z$. In the specific limit discussed above, the ultraviolet asymptotics is thus reduced by that factor.
For the above argument it was important that both $q'_0$ and $q'_y$ are large. The $x$-component did not matter. If only $q'_0$ is assumed to be large, the $k_0$-integration does extend over a large interval, yielding a factor $\Lam^z$, but only as long as $q'_y$ is of the order of the other $y$-components. One thus obtains an effective restriction of $q'_y$ as discussed already for the three-point loop in Sec.~IV.

In summary, we have shown that a fermion loop traversed by a {\em single large}\/ momentum is suppressed by cancellations in the cyclic sum in Eq.~(\ref{INred}).
No such cancellation occurs if three or more momenta $q_i$ are large. In that case three momenta $p_i$, $p_j$, $p_l$ can be far apart from each other so that no systematic cancellation occurs in the cyclic sum.


\subsection{Symmetrized N-point loop with one or two small momenta}

We now show that systematic cancellations occur in the {\em symmetrized}\/ N-point loop in the limit of one or two {\em vanishing}\/ momenta.
This generally reduces the degree of divergence of Feynman diagrams with bosonic external legs coupled to a fermion loop.

We first show that the symmetrized $N$-point loop vanishes, if one of the external momenta vanishes.
The symmetrized $N$-point loop is given by a sum over all permutations of external momenta.
For a vanishing external momentum two fermion lines in the loop carry the same internal momentum. By permutations the vertex with the vanishing momentum is cycled around the loop, yielding the sum
\begin{equation}
 \int \frac{dk_0}{2\pi} \int \frac{d^2\bk}{(2\pi)^2} \,
 \sum_{j=1}^{N-1} G(k-p_j) \prod_{i=1}^{N-1} G(k-p_i) \, .
\end{equation}
Since the denominator of $G(k-p_i)$ is linear in $k_x$, the integrand can be written as a $k_x$-derivative,
$- \frac{\partial}{\partial k_x} \prod_{i=1}^{N-1} G(k-p_i)$.
Performing the $k_x$-integration one thus finds that the symmetrized $N$-point loop vanishes if one leg has a vanishing momentum.
By dimensional analysis, the UV scaling of the symmetrized loop with one fixed external momentum is thus reduced by a factor $\Lambda^{-1}$.

Using analyticity and the invariance under $q_i \mapsto -q_i$ one can conclude that symmetrized loops vanish even quadratically, if a vanishing momentum $q$ enters and leaves the same loop at two distinct vertices, provided that the other momenta remain finite. Due to momentum conservation this is possible only for $N \geq 4$. Hence, the UV scaling of symmetrized loops for $N-2$ large momenta and two fixed external momenta $q$ and $-q$ is reduced by a factor $\Lambda^{-2}$.


\section{Improved power-counting}

We now use the above results on the asymptotic behavior of fermion loops to obtain improved power-counting estimates for the ultraviolet behavior of several classes of Feynman diagrams. We consider contributions to the boson and fermion self-energies, the fermion-boson vertex, and the three-boson vertex.

\subsection{Boson self-energy}

Perturbative contributions to the boson self-energy $\Sg_b(q)$ with $L$ loop-integrations contain $L-1$ boson propagators and $2L$ fermion propagators.
The boson propagators decay as $\Lam^{1-z}$ for large momenta, the fermion propagators as $\Lam^{-2}$. Each loop integration contributes a power $\Lam^{3+z}$. Hence, according to naive power-counting, $L$-loop contributions to the boson self-energy may diverge as $\Lam^{z-1}$. In particular, for $z=3$ a quadratic UV divergence seems possible for any loop order $L$.

However, systematic cancellations of UV divergences occur between distinct contributions (represented by distinct Feynman diagrams) at any given loop order. Feynman diagrams contributing to the boson self-energy $\Sg_b(q)$ contain fermion loops but no open fermion lines. Hence, the external boson legs with ingoing and outgoing momentum $q$ connect directly to fermion loops. The sum over all diagrams of a certain order can be written in terms of symmetrized fermion loops.
In Sec.~V.C we have shown that in symmetrized fermion loops with a fixed (relatively small) momentum $q$, the UV divergences are suppressed by a factor $\Lam^{-1}$ upon symmetrization, and in symmetrized loops with two fixed momenta $q$ and $-q$ by a factor $\Lam^{-2}$.
In Feynman diagrams contributing to the boson self-energy the external momenta $q$ and $-q$ are either attached to two distinct loops or to just one loop.
In both cases there is a power-counting gain of the order $\Lam^{-2}$.
Hence, the sum over all $L$-loop contributions to the boson self-energy behaves as $\Lam^{z-3}$ for large $\Lam$. For $z<3$ this means that the sum is UV-convergent. For $z=3$, quadratic and linear UV divergences must cancel. This cancellation is also imposed by a Ward identity following from current conservation~\cite{metlitski10}.
However, logarithmic UV divergences are not excluded for $z=3$.

At three-loop order, the only candidates for a logarithmic UV divergence for $z=3$ are the Aslamasov-Larkin diagrams (see Fig.~2). However, as discussed already in Sec.~IV, these diagrams are restricted by a kinematic constraint arising from cancellations in the representation of the three-point fermion loop as a cyclic sum, so that the total contribution is UV-finite for any $z<4$.

As shown in Sec.~V.B, a kinematic constraint from a cancellation of contributions to the $N$-point fermion loops in the UV limit occurs only if only one large momentum traverses the loop.
Contributions to the boson self-energy where this condition is {\em not}\/ satisfied occur at four-loop order. An example is shown in Fig.~4.
\begin{figure}[tb]
\centering
\includegraphics[width=7cm]{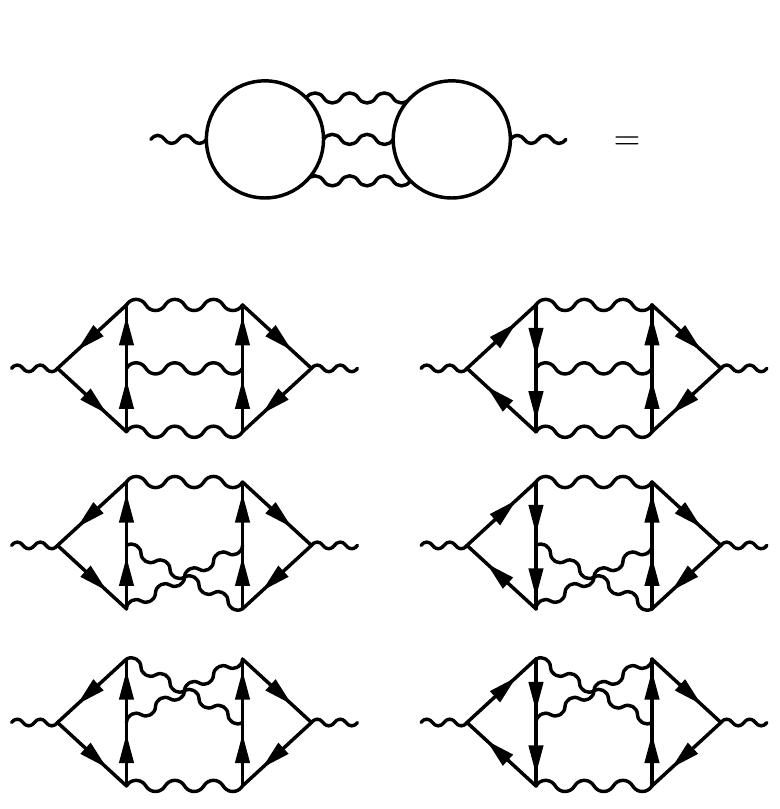}
\caption{A class of four-loop diagrams with fermion loops connected to three boson propagators with possibly large momenta.}
\end{figure}
In Ref.~\onlinecite{holder15}, those contributions (with symmetrized fermion loops) were computed explicitly for $z=3$ in the static limit $q_0 = 0$, and were indeed found to diverge logarithmically. Surprisingly, the divergence turned out to be of the order $\log^5(\Lam/|q_y|)$ instead of the expected simple logarithm. Such a divergence is not renormalizable. The implications of that singularity are not clear at the moment.

It is remarkable that the one-loop result receives qualitative corrections only at four-loop order, while the dynamical exponent $z=3$ remains unchanged at two- and three loop order. Such a situation is peculiar but not unprecedented in quantum field theory.
For example, in non-linear $\sg$-models describing the critical behavior of the Anderson localization transition, only the four-loop contributions lift a degeneracy between distinct symmetry groups and correct the one-loop result for critical localization length and conductivity exponents~\cite{wegner89}.

\subsection{Fermion self-energy}

Perturbative contributions to the fermion self-energy $\Sg_f(k)$ with $L$ loops contain $L$ boson propagators and $2L-1$ fermion propagators. Hence, $L$ loop contributions to the fermion self-energy diverge as $\Lam^2$ in the ultraviolet limit for any $L$ and $z$. This divergent contribution is however momentum and frequency independent and can be absorbed by a shift of the Fermi surface. Physically more interesting is the momentum and frequency dependence of the self-energy. In particular, the derivative $\partial\Sg_f/\partial k_+$ with $k+ = k_x + k_y^2$ determines the anomalous dimension of the fermion fields~\cite{metlitski10}. Since $k_x$ and $k_y^2$ scale as $\Lam^2$, that derivative scales as $\Lam^0$ according to the above power-counting. Hence, logarithmic UV divergences are expected, and indeed occur already at three-loop order (see Fig.~5), as discovered for $z=3$ by Metlitski and Sachdev~\cite{metlitski10}.
The three-point fermion loop in these three-loop contributions is not reduced by any of the cancellations we discussed above, since all three boson momenta can be large.

\begin{figure}[htb]
\centering
\includegraphics[width=8cm]{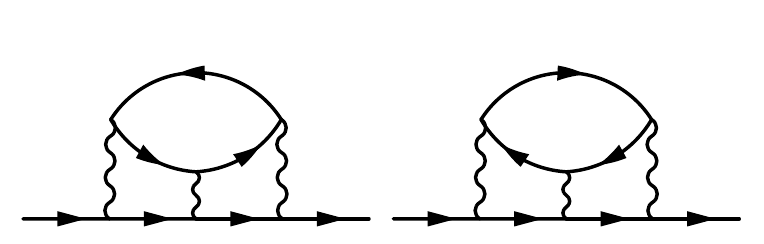}
\caption{Logarithmically divergent three-loop contributions to the fermion self-energy.}
\end{figure}

\subsection{Fermion-boson vertex}

We now discuss the fermion-boson vertex $\Gam(k,q)$ with one boson and two fermion legs. Perturbative $L$-loop vertex corrections contain $L$ boson propagators and $2L$ fermion propagators. Hence, $\Gam(k,q)$ scales as $\Lam^0$ in the UV limit, that is, logarithmic divergences are expected.
The vertex corrections can be grouped in two classes. In the first case, the external boson leg couples to an open fermion line, and in the second to a fermion loop. In Fig.~6 we provide a three-loop example for each case.
\begin{figure}[htb]
\centering
\includegraphics[width=8cm]{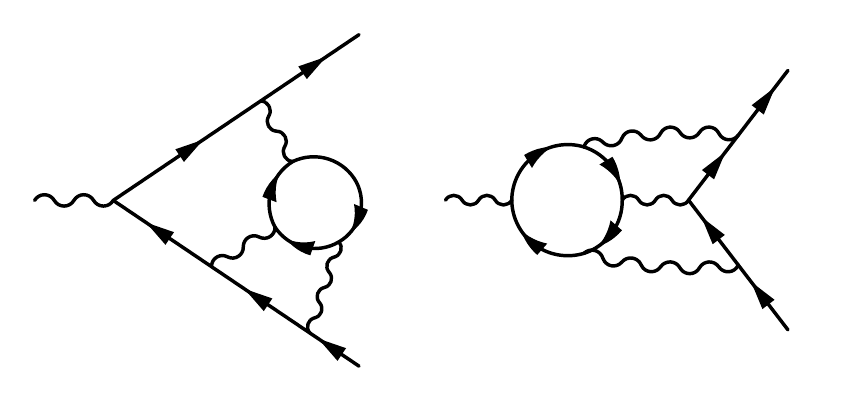}
\caption{Two types of vertex corrections with the external boson leg linked to an open fermion line (left) and to a fermion loop (right).}
\end{figure}
In the first case we do not expect any cancellations, but in the second we have a fermion loop with a fixed boson momentum $q$. Hence, summing all diagrams corresponding to the symmetrized fermion loop, the UV contributions are suppressed by a factor $\Lam^{-1}$ so that the sum is guaranteed to be finite.

The fermion-boson vertex is related to the fermion self-energy by the usual Ward-identity following from charge conservation~\cite{metlitski10}.
Since the fermion self-energy exhibits a logarithmic divergence already at three-loop order, the fermion-boson vertex has to diverge, too. The above argument shows that the divergent contributions to the vertex come exclusively from those diagrams where the boson leg couples to an open fermion line.

\subsection{Three-boson vertex}

As a final example, we consider the three-boson vertex.
Its lowest order (one loop) contribution is simply the symmetrized three-point fermion loop $\Pi_3^{\rm sym}$, which does not exhibit any UV divergence.
$L$-loop corrections to the three-boson vertex contain $L-1$ boson propagators and $2L+1$ fermion propagators. Naive power-counting thus yields a UV scaling of the form $\Lam^{z-3}$. For $z=3$ this is marginal.
However, each of the external boson legs is linked directly to a fermion loop
(see, for example, Fig.~7). Upon loop symmetrization possible UV divergences are thus suppressed. Without investigating the exact degree of suppression we can say that in the most interesting physical case $z=3$ all corrections will be UV finite.
The UV convergence of $N$-loop vertices with $N>3$ is also improved by cancellations in the symmetrized fermion loops, but they are already UV convergent even within naive power-counting for $z < 2N-3$.
\begin{figure}[htb]
\centering
\includegraphics[width=5cm]{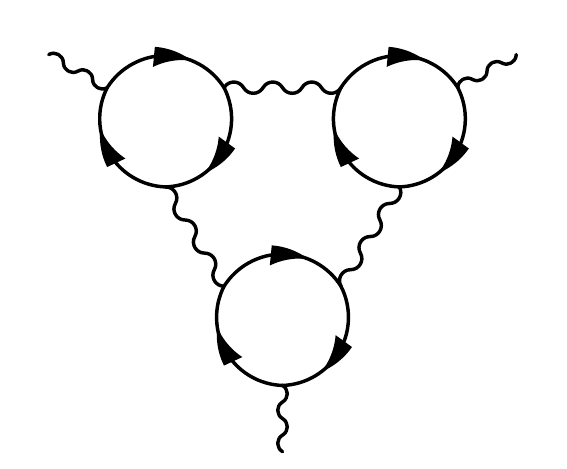}
\caption{Four-loop diagram contributing to the three-boson vertex.}
\end{figure}
%


\section{Conclusion}

We have analyzed general properties of the perturbative loop expansion for two-dimensional quantum critical metals with singular forward scattering.
Important cases falling in this class of systems are metals at an Ising nematic QCP and metals coupled to a $U(1)$ gauge field. Our analysis is based on the effective field theory for such systems~\cite{lee09,metlitski10}, extended to arbitrary dynamical exponents $z>2$ as proposed earlier by Nayak and Wilczek~\cite{nayak94}.

We have derived asymptotic properties of fermion loops appearing as subdiagrams of Feynman diagrams contributing to the perturbation series.
Substantial cancellations occur when summing over contributions corresponding
to permutations of vertices at fermion loops with a small external momentum $q$ or two small momenta $q$ and $-q$ (while all the other momenta become large).
Furthermore, fermion loops are suppressed below the naive power-counting estimate when they are traversed by only one large bosonic momentum.

Using these properties we have shown in a number of specific examples how one can perform a {\em sharp}\/ ultraviolet power-counting that takes all possible cancellations into account.
Most importantly, for the boson self-energy there is always a gain of order
$\Lam^2$. As a consequence, perturbative contributions to that quantity are UV convergent for $z<3$. Hence, perturbative self-energy corrections to the boson propagator are finite at all loop orders for $z<3$. In the static limit $q_0 = 0$, the boson self-energy is then fixed by dimensional analysis to be proportional to $q_y^{z-1}$ with a finite prefactor.
A bare $z<3$ is thus not renormalized by fluctuations.
On the other hand, logarithmic divergences are possible in the important physical case with a dynamical exponent $z=3$, and indeed do occur at four-loop order~\cite{holder15}.
For $z>3$ there are UV divergences proportional to $\Lam^{z-3}$.

Obviously, $z=3$ is a critical value at the boundary between the qualitatively different cases $z<3$ and $z>3$. Interestingly, the same critical value $z=3$ separates two qualitatively different behaviors of the {\em compressibility}, if the chemical potential $\mu$ couples to the bosonic mass appearing away from the QCP. For $z < 3$ the compressibility is finite, while for $z \geq 3$ it diverges upon approaching the QCP~\cite{metlitski10}. Our result that a bare $z < 3$ is not renormalized by fluctuations at any loop order thus implies that the compressibility remains finite in that case.

There are no systematic cancellations reducing the UV divergences of the fermion self-energy. Hence, the fermionic field renormalization obtained from a first momentum derivative of the self-energy diverges logarithmically for any $z$. In contrast to the behavior of the boson self-energy, the degree of divergence of the fermion self-energy is independent of $z$. Logarithmic divergences obtained in perturbation theory can be summed up to yield anomalous power-law scaling, as demonstrated in the three-loop calculation for $z=3$ by Metlitski and Sachdev~\cite{metlitski10}.

At this point, the perturbative singularity structure of the above class of critical metals seems to be clarified for $z<3$. The one-loop fixed point is only modified by an anomalous scaling of the fermion fields. However, the final fate of the theory in the special but physically relevant case $z=3$ remains open.
Organizing the loop expansion in terms of symmetrized fermion loops and exploiting the cancellations described above should be useful for any future work on fluctuation corrections. In particular, a numerical computation of fluctuation corrections is greatly facilitated by incorporating the cancellations directly via symmetrized loops.


\begin{acknowledgments}
We are grateful to A.~Eberlein, M.~Metlitski, D.~Vollhardt, and R.~Zeyher for valuable discussions.
\end{acknowledgments}

\newpage


\begin{appendix}


\section{Integration of loops by residues}

Here we derive Eq.~(\ref{IN}) for the $N$-point fermion loop
\begin{equation}
 I_N(p_1,\dots,p_N) =
 \int_{-\infty}^{\infty} \! \frac{dk_0}{2\pi} \! 
 \int_{-\infty}^{\infty} \! \frac{dk_x}{2\pi} \!
 \int_{-\infty}^{\infty} \! \frac{dk_y}{2\pi} \!
 \prod_{i=1}^N G(k-p_i) \, ,
\end{equation}
where $G(k) = \left[ k_x + k_y^2 - i\{k_0\} \right]^{-1}$
with $\{k_0\} = k_0/|k_0|^{\alf}$.
The $k_x$-integration can be performed by closing the integration contour in the upper complex half-plane and using the residue theorem. There are $N$ simple poles situated at $k_x = i\{k_0 - p_{i0}\} + p_{ix} - (k_y - p_{iy})^2$ with $i=1,\dots,N$.
Summing the contributions from all poles in the upper half plane ($k_0 > p_{i0}$) yields
\begin{widetext}
\begin{eqnarray}
 I_N(p_1,\dots,p_N) &=& i\int_{-\infty}^{\infty} \frac{dk_0}{2\pi}
 \int_{-\infty}^{\infty} \frac{dk_y}{2\pi} \, 
 \sum_{i=1}^N \Theta(k_0 - p_{i0}) \nonumber \\
 && \hskip -1cm \times \prod_{j \neq i} \, \frac{1}
 {i\{k_0 - p_{i0}\} - i\{k_0 - p_{j0}\} + p_{ix} - p_{jx} +
 2(p_{iy} - p_{jy})k_y - p_{iy}^2 + p_{jy}^2} \, . \hskip 1cm
\end{eqnarray}
Now the $k_y$-integration can be performed analogously.
For each $i$ there are $N-1$ simple poles at
\begin{equation}
 k_y = \frac{i\{k_0 - p_{j0}\} - i\{k_0 - p_{i0}\} - p_{ix} + p_{jx} +
 p_{iy}^2 - p_{jy}^2}{2(p_{iy} - p_{jy})} \nonumber
\end{equation}
with $j \neq i$. These poles are situated in the upper complex half plane if (and only if) $\frac{p_{i0} - p_{j0}}{p_{iy} - p_{jy}} > 0$. Closing the integration contour in the upper complex half plane thus yields
\begin{eqnarray} \label{A3}
 I_N(p_1,\dots,p_N) &=& 
 - \int_{-\infty}^{\infty} \frac{dk_0}{2\pi}
 \sum_{i} \sum_{j\neq i} \Theta(k_0 - p_{i0}) 
 \Theta\Big(\frac{p_{i0} - p_{j0}}{p_{iy} - p_{jy}}\Big)
 \frac{1}{2(p_{iy} - p_{jy})} \nonumber \\
 &\times& \prod_{l \neq i,j} \Big[
 i\{k_0 - p_{i0}\} - i\{k_0 - p_{l0}\} + p_{ix} - p_{lx} - p_{iy}^2 + p_{ly}^2
 \nonumber \\
 &+& \frac{p_{iy} - p_{ly}}{p_{iy} - p_{jy}} (i\{k_0 - p_{j0}\}
 - i\{k_0 - p_{i0}\} - p_{ix} + p_{jx} + p_{iy}^2 - p_{jy}^2) \Big]^{-1}.
 \hskip 5mm
\end{eqnarray}
Defining
\begin{eqnarray}
 D_{ijl} &=& - \det \left( \begin{array}{cc}
 p_{ix} - p_{jx} & p_{ix} - p_{lx} \\
 p_{iy} - p_{jy} & p_{iy} - p_{ly} \end{array} \right) = p_{ix} (p_{ly} - p_{jy}) + {\rm cyc} \, , \\[2mm]
 F_{ijl} &=& (p_{jy} - p_{iy})(p_{ly} - p_{jy})(p_{iy} - p_{ly}) 
 \, , \\[2mm]
 \Om_{ijl}(k_0) &=& \{k_0 - p_{i0}\} (p_{ly} - p_{jy}) + {\rm cyc} \, ,
\end{eqnarray}
Eq.~(\ref{A3}) can be written as
\begin{eqnarray}
 I_N(p_1,\dots,p_N) &=& 
 - \int_{-\infty}^{\infty} \frac{dk_0}{4\pi}
 \sum_{i} \sum_{j\neq i} \Theta(k_0 - p_{i0})  \Theta\Big(\frac{p_{i0} - p_{j0}}{p_{iy} - p_{jy}}\Big)
 (p_{iy} - p_{jy})^{N-3}  \prod_{l \neq i,j}
 \left[ D_{ijl} + F_{ijl} + i\Om_{ijl}(k_0) \right]^{-1} . \hskip 8mm
 \end{eqnarray}
Defining
$J_{ijl}(k_0) = \left[ D_{ijl} + F_{ijl} + i\Om_{ijl}(k_0) \right]^{-1}$,
and using the antisymmetry $J_{ijl}(k_0) = - J_{jil}(k_0)$, one obtains $I_N(p_1,\dots,p_N)$ as a sum over indices $i$ and $j$ restricted to $i<j$ in the form Eq.~(\ref{IN}).


\section{Derivation of reduction formula}

In this Appendix we derive the expression Eq.~(\ref{INred}) for the $N$-point loop from Eq.~(\ref{IN}).
The crucial step is the identity
\begin{eqnarray} \label{redN}
 (p_{iy} - p_{jy})^{N-3} \prod_{l \neq i,j} J_{ijl}(k_0) &=&
 \sum_{l \neq i,j}
 \Big[ \prod_{\nu \neq i,j,l} f_{ijl,\nu}^{-1}(k_0) \Big]  J_{ijl}(k_0) \, ,
\end{eqnarray}
where $f_{ijl,\nu}(k_0)$ is defined as in Eq.~(\ref{fk0}), that is,
\begin{equation}
 f_{ijl,\nu}(k_0) = 
 \frac{D_{ijl} J_{ij\nu}^{-1}(k_0) - D_{ij\nu} J_{ijl}^{-1}(k_0)}
 {D_{ijl} (p_{iy} - p_{jy})} \, .
\end{equation}
This can be verified by first considering the cases $N=4$ and $N=5$, and then proceeding by induction.
For $N=4$, Eq.~(\ref{redN}) reads (suppressing the $k_0$-dependence)
\begin{equation} \label{red4}
 (p_{iy} - p_{jy}) J_{ijl_1} J_{ijl_2} =
 f_{ijl_1,l_2}^{-1} J_{ijl_1} + f_{ijl_2,l_1}^{-1} J_{ijl_2} \, ,
\end{equation}
which follows from
\begin{eqnarray}
 J_{ijl_1} J_{ijl_2} &=& J_{ijl_1} J_{ijl_2} \Big(
 \frac{D_{ijl_1} J_{ijl_1}}{D_{ijl_1} J_{ijl_1} - D_{ijl_2} J_{ijl_2}} +
 l_1 \lra l_2 \Big)
 =
 \frac{D_{ijl_1}}{D_{ijl_1} J_{ijl_2}^{-1} - D_{ijl_2} J_{ijl_1}^{-1}} J_{ijl_1} + 
 l_1 \lra l_2 \, .  \nonumber
\end{eqnarray}
For $N=5$, one finds from repeated application of Eq.~(\ref{red4})
\begin{eqnarray}
 (p_{iy} - p_{jy})^2 J_{ijl_1} J_{ijl_2} J_{ijl_3} &=&
 f_{ijl_1,l_2}^{-1} f_{ijl_1,l_3}^{-1} J_{ijl_1} 
+ f_{ijl_2,l_1}^{-1} f_{ijl_2,l_3}^{-1} J_{ijl_2}
   + (f_{ijl_2,l_1}^{-1} f_{ijl_3,l_2}^{-1} + 
      f_{ijl_1,l_2}^{-1} f_{ijl_3,l_1}^{-1}) J_{ijl_3} \nonumber \\
&=& f_{ijl_1,l_2}^{-1} f_{ijl_1,l_3}^{-1} J_{ijl_1} +
   f_{ijl_2,l_1}^{-1} f_{ijl_2,l_3}^{-1} J_{ijl_2} 
+ f_{ijl_3,l_1}^{-1} f_{ijl_3,l_2}^{-1} J_{ijl_3} ,
\end{eqnarray}
where in the last step we have used
$f_{ijl_2,l_1}^{-1} f_{ijl_3,l_2}^{-1} + f_{ijl_1,l_2}^{-1} f_{ijl_3,l_1}^{-1} =
 f_{ijl_3,l_1}^{-1} f_{ijl_3,l_2}^{-1}$, which can be verified by explicit calculation.
It is now clear that Eq.~(\ref{redN}) follows from $(N-3)$-fold iterated application of Eq.~(\ref{red4}).

Inserting Eq.~(\ref{redN}) into Eq.~(\ref{IN}) yields
\begin{eqnarray}
 I_N(p_1,\dots,p_N) &=& \sum_{i<j} \sum_{l \neq i,j}
 \int_{p_{j0}}^{p_{i0}} \frac{dk_0}{4\pi} \,
 \Theta\Big( \frac{p_{i0} - p_{j0}}{p_{iy} - p_{jy}} \Big) 
 \Big[ \prod_{\nu \neq i,j,l} f_{ijl,\nu}^{-1}(k_0) \Big]
 J_{ijl}(k_0) \, .
\end{eqnarray}
Using the invariance of $J_{ijl}(k_0)$ and $f_{ijl,\nu}(k_0)$ under cyclic permutations of $i,j,l$ one obtains $I_N(p_1,\dots,p_N)$ in the cyclic form Eq.~(\ref{INred}).

\end{widetext}

\end{appendix}



\end{document}